\documentclass[preprint]{aastex62}
\shorttitle{General circulation collapses in shortwave absorptive atmosphere}
\shortauthors{Kang and Wordsworth}
\begin{document}
\title{Collapse of the general circulation in shortwave-absorbing atmospheres: an idealized model study}

\correspondingauthor{Wanying Kang}
\email{wanyingkang@g.harvard.edu}

\author[0000-0002-4615-3702]{Wanying Kang}
\affil{School of Engineering and Applied Sciences\\
  Harvard University\\
  Cambridge, MA 02138, USA}

\author{Robin Wordsworth}
\affiliation{School of Engineering and Applied Sciences\\
  Harvard University\\
  Cambridge, MA 02138, USA}

\begin{abstract}
 The response of the general circulation in a dry atmosphere to various atmospheric shortwave absorptivities is investigated in a three-dimensional general circulation model with grey radiation. Shortwave absorption in the atmosphere reduces the incoming radiation reaching the surface but warms the upper atmosphere, significantly shifting the habitable zone toward the star. The strong stratification under high shortwave absorptivity suppresses the Hadley cell in a manner that matches previous Hadley cell scalings. General circulation changes may be observable through cloud coverage and superrotation. The equatorial superrotation in the upper atmosphere strengthens with the shortwave opacity, as predicted based on the gradient wind of the radiative-convective equilibrium profile. There is a sudden drop of equatorial superrotation at very low shortwave opacity. This is because the Hadley cell in those cases are strong enough to fill the entire troposphere with zero momentum air from the surface. A diurnal cycle (westward motion of substellar point relative to the planet) leads to acceleration of the equatorial westerlies in general, through the enhancement of the equatorward eddy momentum transport, but the response is not completely monotonic, perhaps due to the resonance of tropical waves and the diurnal forcing.
\end{abstract}

%\keywords{}

\section{Introduction}
\label{sec:intro}

%% Importance of general circulation in general
While single column radiative-convective equilibrium (RCE) models capture a lot of useful features of a climate system, they ignore the horizontal inhomogeneity, an important factor in most climate feedbacks. For example, activation of the ice-albedo feedback is controlled by the lowest rather than the average surface temperature over the globe, and the former is to a large extent affected by the temporal and spatial distribution of insolation, and the horizontal heat transport \cite[e.g,][]{Yang-Boue-Fabrycky-et-al-2014:strong, Linsenmeier-Pascale-Lucarini-2015:climate}. Cloud coverage and the accompanying radiative effects can change completely when the general circulation changes \cite[e.g,][]{Wang-Liu-Tian-et-al-2016:effects, Kang-2019:mechanisms}. Given the configuration of a planet, one needs to understand the general circulation to predict the climate state and habitability. In turn, general circulation features, if detectable, can provide extra information to constrain the planet's climate.

%% Relevance of a shortwave absorptive atmosphere: Mars, exoplanets around red dwarfs
In this work, we investigate the atmospheric general circulation under different shortwave absorptivities. Shortwave absorption at visible wavelengths by the atmosphere is weak on Earth. However, ultraviolet absorption by ozone gives rise to Earth's strong temperature inversion in the stratosphere and shapes the stratospheric circulation \cite[][]{Andrews-Holton-Leovy-1987:middle}. The magnitude of shortwave absorption can be affected by the solar radiation spectrum, the atmosphere composition, and the presence of aerosols, including dust and clouds. As the radiation spectrum of M-dwarfs is shifted to the infrared \cite[][]{Phillips-1994:physics}, H$_2$O, CO$_2$ and other greenhouse gases can absorb the incoming radiation more efficiently \cite[][]{Kasting-Whitmire-Reynolds-1993:habitable, Wordsworth-Pierrehumbert-2013:water}. Higher concentrations of ozone due to either biological or abiotic \cite[][]{Wordsworth-Pierrehumbert-2014:abiotic, Luger-Barnes-2015:extreme} oxygen production can also enhance the shortwave absorption from sun-like stars. Dust/cloud layers can also absorb shortwave radiation if the particles have a large imaginary refractive index. The dust storms on Mars, for example, lead to strong shortwave absorption in the atmosphere, which in turn helps to maintain the storms once they become large enough \cite[][]{Conrath-1975:thermal, Haberle-1986:interannual, Rafkin-2009:positive}.

%% Hypothesis: stronger stratification may suppress circulation, leaving stronger meridional temperature gradient 
The atmospheric thermal structure with different shortwave absorptivities has been studied in \cite{Kurokawa-Nakamoto-2012:effects} using a 1D RCE model. \cite{Hansen-2008:absorption, Guillot-2010:radiative, Heng-Hayek-Pont-et-al-2012:effects} derived analytical or semi-analytical solutions of the radiative-equilibrium temperature profile and of the outgoing radiation for hot Jupiters with an internal heat source. However, all of these studies only considered the vertical direction. The general circulation response has not been investigated before systematically in 3D. Here, using a 3D GCM with semi-grey radiation scheme, we study the general circulation of a dry atmosphere on Earth-like planets, and the potential observable consequences, over a full range of shortwave absorptivities. The effects of changing the planet's rotation rate and the length of the diurnal cycle are also discussed. Thanks to the highly idealized model setup, analytic explanations can be provided for the two main circulation responses to an enhanced shortwave absorption: the suppression of the meridional overturning circulation, and the enhancement of the equatorial superrotation in the upper atmosphere.

\section{Methods}
\label{sec:methods}
%% Isca model intro

We use Isca v1.0, a framework for idealized modelling of the general circulation of planetary atmospheres. It is developed by \cite{Vallis-Colyer-Geen-et-al-2018:isca} on top of the Flexible Modeling System (https: //www.gfdl.noaa.gov/fms/, Geophysical Fluid Dynamics Laboratory) to include various forcing and radiation choices with different levels of complexity. In order to compare with the analytic solution under the grey-atmosphere assumption, we choose a two-band grey-atmosphere radiation scheme as used in \cite{Frierson-Held-Zurita-Gotor-2006:gray}, and change the shortwave/longwave optical depths. In all experiments, the insolation is set to 1360 W/m$^2$, orbital obliquity is zero and the surface pressure is $10^5$~Pa. The hydrological cycle is turned off by setting atmospheric humidity to zero initially and suppressing surface evaporation during the simulation. Dry convection is allowed, in which the vertical temperature profile is restored to a dry adiabat with a time scale of 1800~s. The atmosphere is coupled to a mixed layer with 30-m water heat capacity and zero albedo. Surface friction is parameterized by a relaxation to zero speed with a restoring strength of $5\times10^{-5}$ s$^{-1}$. We choose a relatively low resolution, T21 for horizontal grid and 25 evenly distributed layers in $z$, in order to run enough simulations for long enough to study how the general circulation and equatorial jet responds to different parameters. We test the sensitivity to the resolution by running two groups of simulations at a higher resolution, T42. Each simulation is run for 50 years in total, and climatology is calculated from the last 20 years. We have verified that equilibrium state has been reached by the 30\textsuperscript{th} year, with no observable secular trend in either the zonal wind or the temperature distribution.

%% Experiment design
To investigate how the general circulation changes with the shortwave absorptivity of the atmosphere, we run Isca at eight different shortwave optical depths $\tau_\infty^s$, varying from 0 (transparent to shortwave) to 12 (very opaque), while fixing the longwave optical depth $\tau^\infty$ (i.e., greenhouse effect) at 4. The rotation rate, as another factor that significantly affects the general circulation, is also explored. Eight $\tau_\infty^s$-varying experiments are run under 0.25 and 1 times Earth's Coriolis coefficient, respectively, to show that the scaling of the general circulation suppression is not sensitive to cover a larger range of exoplanets and to test current existing scalings for the Hadley cell. We do not simulate the cases with even faster rotation rate than Earth, because those planets are usually far away from the star and thus are difficult to observe due to their long orbital period, and also because it takes much higher horizontal resolution to capture the dynamics on those planets \cite[][]{Kaspi-Showman-2015:atmospheric}. Finally, to sketch a continuous picture of how the circulation and equatorial superrotation characteristics vary with the strength of the diurnal cycle, we make the length of a day to be 4 times longer and shorter, and also try to force the model with daily-mean insolation, while keeping the Coriolis coefficient at the standard Earth value. The experiment setups are summarized in Table.~\ref{tab:experiment-setup}. Given the highly idealized model setup we use here, we do not expect our model to correctly capture all general circulation characteristics, especially in presence of condensible components, but it may provide useful implications for exoplanet atmospheres without condensible components. 

\begin{table*}[htp!]
\centering
\caption{Experiment setups} \label{tab:experiment-setup}
\begin{tabular}{l|cccc}
\tablewidth{0pt}
\hline
\hline
Experiment Name          & Coriolis coef. (times earth value) & Diurnal cycle & Day length (1 day) & Resolution \\
\hline
\decimals
diurnal                  & 1                                 & yes           & 1                  & T21        \\
%diurnal, $\Omega\times4$ & 4                                 & yes           & 1                  & T21        \\
diurnal, $\Omega/4$      & 0.25                              & yes           & 1                  & T21        \\
daily mean               & 1                                 & no            & --                 & T21        \\
%daily mean, $\Omega\times4$ & 4                              & no            & --                 & T21        \\
daily mean, $\Omega/4$   & 0.25                              & no            & --                 & T21        \\
  diurnal fast           & 1                                 & yes           & 0.25               & T21        \\
  diurnal slow           & 1                                 & yes           & 4                  & T21        \\
  daily, high-res        & 1                                 & no            & --                 & T42        \\
  daily, $\Omega/4$, high-res & 0.25                         & no            & --                 & T42        \\
\hline
\end{tabular}
\end{table*}

\section{Results}
\label{sec:results}

\subsection{Radiative-convective equilibrium temperature profile on planets with a solar-absorptive atmosphere}
\label{sec:RCE-analytical}

We first sketch the derivation of the radiative-convective equilibrium (RCE) temperature for an idealized gray atmosphere, which absorbs not only thermal radiation but also solar radiation, following Section~4.3.4-4.3.5 of \citet{Pierrehumbert-2010:principles}. In reality, this shortwave absorption may happen on planets with a highly oxidized atmosphere, such that a thick ozone layer could form in the upper atmosphere, or alternatively on planets with a permanent hydrocarbon haze layer. Significant absorption of solar radiation may also happen on planets surround M-stars, as the solar radiation from M-star peaks in the infrared frequency where CO$_2$, H$_2$O and other common greenhouse gases have strong absorption. We assume that the atmosphere has uniform optical depth versus frequency for thermal radiation (gray atmosphere), $\tau_\infty$, and for solar radiation $\tau_\infty^s$. The ratio between the two optical depth are denoted as $\chi$,
$$\chi\equiv\tau_\infty/\tau_\infty^s.$$
If the surface has zero albedo and the solar zenith angle equals the mean thermal emission angle, the two-stream radiative transfer equations can be written
\begin{eqnarray}
  \frac{dF_{\uparrow}}{d\tau}&=&F_{\uparrow}-\sigma T(\tau)^4\label{eq:trans-I+}\\
  \frac{dF_{\downarrow}}{d\tau}&=&-F_{\downarrow}+\sigma T(\tau)^4\label{eq:trans-I-}\\
  \frac{dS_{\downarrow}}{d\tau}&=&-S_{\downarrow}/\chi.\label{eq:trans-S-}
\end{eqnarray}
Here $\tau$ is the optical depth for thermal radiation counted from the top of the atmosphere (TOA), $F$ is thermal radiation flux, $S$ is solar radiation flux, subscript $\uparrow$ denotes upward fluxes, and $\downarrow$ denotes downward fluxes. The radiative equilibrium temperature is achieved when the net heating rate vanishes at each level,
\begin{eqnarray}
  \frac{d}{d\tau}(F_{\uparrow}-F_{\downarrow}-S_{\downarrow})&=&0. \label{eq:heating-rate}
\end{eqnarray}

We integrate Eq~(\ref{eq:heating-rate}) and evaluate it at TOA to get
\begin{eqnarray}
  F_{\uparrow}(\tau)-F_{\downarrow}(\tau)&=&S_{\downarrow}(\tau)=S_{\downarrow}(0)e^{-\tau/\chi}.\label{eq:Idiff}
\end{eqnarray}

Adding Eq~(\ref{eq:trans-I+}) and Eq~(\ref{eq:trans-I-}) together then yields
\begin{eqnarray}
  \frac{d(F_{\uparrow}+F_{\downarrow})}{d\tau}&=&F_{\uparrow}-F_{\downarrow}=S_{\downarrow}(0)e^{-\tau/\chi}\nonumber\\
  F_{\uparrow}(\tau)+F_{\downarrow}(\tau)&=&F_{\uparrow}(0)+F_{\downarrow}(0)+S_{\downarrow}(0)\chi\left(1-e^{-\tau/\chi}\right)\nonumber\\
  &=&S_{\downarrow}(0)+S_{\downarrow}(0)\chi\left(1-e^{-\tau/\chi}\right).\label{eq:Isum}
\end{eqnarray}

In the above equation, we have used $F_{\downarrow}(0)=0$, i.e., there is no downward thermal radiation flux at TOA. Substituting Eq~(\ref{eq:trans-I+})-(\ref{eq:trans-S-}) into Eq~(\ref{eq:heating-rate}) we then get
\begin{eqnarray}
 % 2\sigma T(\tau)^4&=&F_{\uparrow}(\tau)+F_{\downarrow}(\tau)+S_{\downarrow}(\tau)/\chi=S_{\downarrow}(0)\left[1+\chi-(\chi-\chi^{-1})e^{-\tau/\chi}\right]\nonumber\\
  T(\tau)&=&T_{\mathrm{skin}}\left[1+\chi-(\chi-\chi^{-1})e^{-\tau/\chi}\right]^{1/4},\label{eq:Ta-tau}
\end{eqnarray}
where $T_{\mathrm{skin}}=(S_{\downarrow}(0)/2\sigma)^{1/4}$ is the skin temperature. One can see that the radiative equilibrium temperature could either increase or decrease with height, depending on the ratio of the optical depth of solar radiation and thermal radiation $\chi$ \citep{Pierrehumbert-2010:principles}. When $\chi>1$ ($\tau_\infty>\tau_\infty^s$), the atmosphere cools with height in radiative equilibrium; while the opposite is true when $\chi<1$ ($\tau_\infty<\tau_\infty^s$), as $\chi-\chi^{-1}$ in Eq~(\ref{eq:Ta-tau}) changes sign. As $\chi\rightarrow 0$, surface temperature asymptotically approaches $T_{\mathrm{skin}}$. This is not surprising, because the only heat source at the surface is from the atmosphere, as the stratosphere. But one should also note that this limit is hard to achieve on real planets, which always have nonzero geothermal heat fluxes.

To see whether convection can be triggered, we need to evaluate the surface temperature. With the help of  Eq~(\ref{eq:Idiff}) and Eq~(\ref{eq:Isum}), the energy balance at the surface gives
\begin{eqnarray}
  \sigma T_g^4&=&S_{\downarrow}(\tau_\infty)+F_{\downarrow}(\tau_\infty)=S_{\downarrow}(0)\left[\frac{1+\chi}{2}+\frac{1-\chi}{2}e^{-\tau_\infty/\chi}\right]\nonumber\\
%              &=&\left[S_{\downarrow}(0)e^{-\tau_\infty/\chi}\right] + \left[S_{\downarrow}(0)\frac{1+\chi}{2}\left(1-e^{-\tau_\infty/\chi}\right)\right]\nonumber\\
              %&=&S_{\downarrow}(0)\left[\frac{1+\chi}{2}+\frac{1-\chi}{2}e^{-\tau_\infty/\chi}\right] \nonumber\\
  T_g&=&T_{\mathrm{skin}}\left[(1+\chi)-(\chi-1)e^{-\tau_\infty/\chi}\right]^{1/4}.\label{eq:Tg}
\end{eqnarray}

$T_g$ does not match the atmospheric temperature right above the surface $T(\tau_\infty)$, and a warmer surface than the bottom atmosphere could trigger convection. Taking the difference between Eq.~(\ref{eq:Tg}) and Eq.~(\ref{eq:Ta-tau}) gives
\begin{eqnarray}
  \label{eq:T-jump}
T_g-T(\tau_\infty)=T_{\mathrm{skin}}\left\{\left[(1+\chi)-(\chi-1)e^{-\tau_\infty/\chi}\right]^{1/4}-\left[1+\chi-(\chi-\chi^{-1})e^{-\tau_\infty/\chi}\right]^{1/4}\right\}.
\end{eqnarray}
When $\chi<1$, $T_g$ is higher than $T(\tau_\infty)$, indicating that the atmosphere is convective unstable right at the surface. When $\chi>1$, $T_g$ is lower than $T(\tau_\infty)$, and convection is suppressed. As $\chi$ keeps increasing, the temperature jump at the surface approaches zero.

We can recover the shortwave transparent limit by letting $\chi\rightarrow \infty$,
\begin{eqnarray}
  T^{\chi\rightarrow\infty}(\tau)&=&\lim_{\chi\rightarrow\infty}T_{\mathrm{skin}}\left[1+\chi-(\chi-\chi^{-1})e^{-\tau/\chi}\right]^{1/4}\nonumber\\
                                 &=&T_{\mathrm{skin}}\left[1+\lim_{\chi\rightarrow\infty}\chi\left(1-e^{-\tau/\chi}\right)\right]^{1/4}\nonumber\\
%&=&T_{\mathrm{skin}}\left[1+\chi\left(1-1+\tau/\chi\right)\right]^{1/4}\nonumber\\
  &=&T_{\mathrm{skin}}\left(1+\tau\right)^{1/4}.\label{eq:Ta-tau-chi-infty}\\
         %&=&T_{\mathrm{skin}}\left[1+\chi\frac{\tau}{\chi}\right]^{1/4}=T_{\mathrm{skin}}\left(1+\tau\right)^{1/4}.\label{eq:Ta-tau-chi-infty}\\
  T_g^{\chi\rightarrow\infty}&=&\lim_{\chi\rightarrow\infty}T_{\mathrm{skin}}\left[(1+\chi)+(1-\chi)e^{-\tau_\infty/\chi}\right]^{1/4}\nonumber\\
                                 &=&\lim_{\chi\rightarrow\infty}T_{\mathrm{skin}}\left[2+\chi\left(1-e^{-\tau_\infty/\chi}\right)\right]^{1/4}\nonumber\\
%&=&\lim_{\chi\rightarrow\infty}T_{\mathrm{skin}}\left[2+\chi\left(1-1+\tau_\infty/\chi\right)\right]^{1/4}\nonumber\\
 &=& T_{\mathrm{skin}}\left(2+\tau_\infty\right)^{1/4}.\label{eq:Tg-chi-infty}
\end{eqnarray}

\begin{figure*}[htp!]
 \centering
 \includegraphics[width=0.9\textwidth]{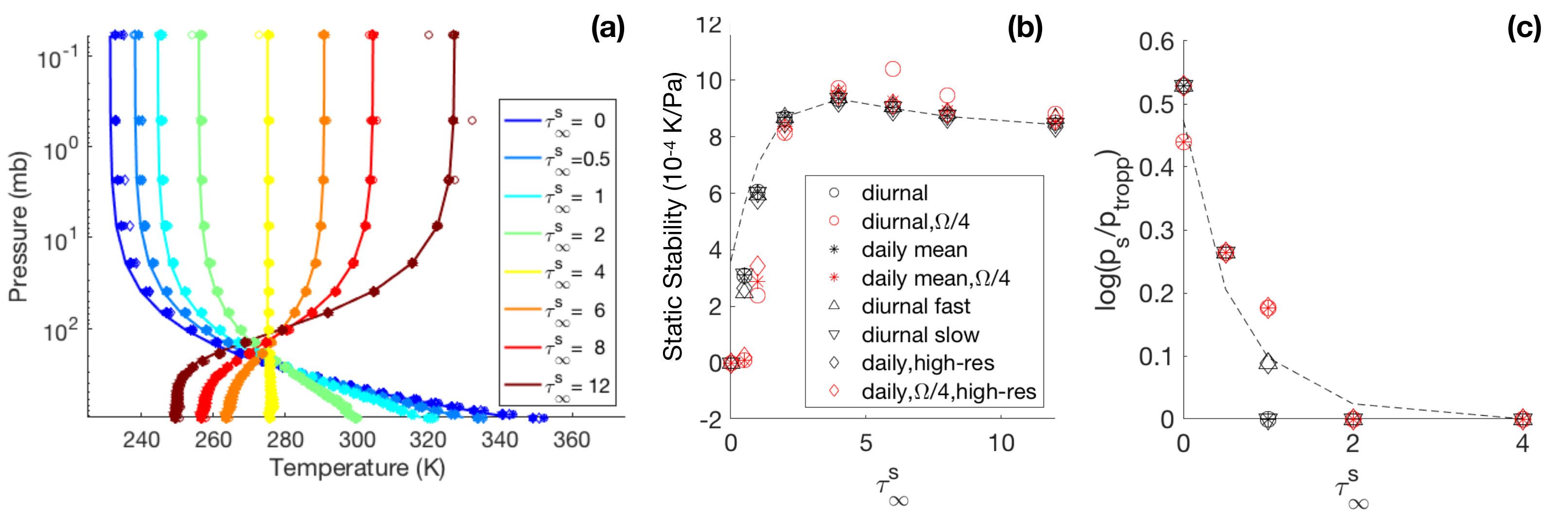}
 \caption{One-dimensional RCE provides a reasonable estimate for the thermal structure in the 3D GCM. Panel (a) shows the vertical temperature profiles in simulations (markers) and in the analytical solution (lines) with various shortwave optical depths marked by different colors. Panel (b) shows the static stability $S_p$ at (0N, 900~mb) as a function of shortwave optical depth. The analytical solution is given by the dashed curve and the GCM results by the various markers. Panel (c) shows the estimated and modeled tropopause height as a function of shortwave optical depth. In both b) and c), the dashed black line is the RCE estimate, while the scatter points show the GCM results. Simulations shown include all 8 configurations, with different Coriolis coefficients and period of diurnal cycle. Each configuration involves 8 experiments with different shortwave absorptivity. The relationship between pressure and optical depth is $p=p_s\frac{\tau}{\tau_\infty}$.}
 \label{fig:S-Ptropp-RCEvsIsca}
\end{figure*}

We show the radiative equilibrium temperature profile from Eq~(\ref{eq:Ta-tau}) and Eq~(\ref{eq:Ta-tau-chi-infty}) in Fig.~\ref{fig:S-Ptropp-RCEvsIsca}(a) for different $\tau_\infty^s$ (lines). It matches well with the temperature vertical profiles at the equator diagnosed from GCM simulations (markers), regardless of the model configuration (varying rotation rate, diurnal cycle etc.).\footnote{The equatorial temperature profiles diagnosed from GCM simulations are slightly colder than predicted due to the adiabatic cooling by upward motion there. The global mean temperature profiles fit even better. We show the equatorial temperature profiles because they are used to calculate the tropical static stability and then the Hadley cell scalings later.} As $\tau_\infty^s$ increases, more and more solar radiation is absorbed in the atmosphere before reaching the surface, leading to surface cooling and upper atmosphere warming. The different vertical thermal structures may be distinguishable in real gas spectrum. The surface cooling can shift the habitable zone toward the host star by a large amount, at least in this idealized setup. See also \cite{Pujol-Fort-2002:effect} for an analytical Komabayashi-Ingersoll limit (K-I limit) when considering both shortwave and longwave absorption by the atmosphere.
%With an infrared window band, the runaway limit even disappears when shortwave absorption is strong enough. 

The static stability of the radiative equilibrium temperature can be obtained by substituting the radiative equilibrium temperature (Eq.~\ref{eq:Ta-tau}) into the definition of static stability $S_p=-T\partial_p\ln\Theta$ ($\Theta$ is potential temperature, defined as $T(p_s/p)^\kappa$, where $\kappa=R/C_p=2/7$ and $p_s$ is surface pressure.), for different $\chi$. This analytical $S_p$ at (0N, 900~mb) is plotted against $\tau_\infty^s$ in black dashed curve in Fig.~\ref{fig:S-Ptropp-RCEvsIsca}(b). $S_p$ at the same location is also diagnosed in the 3D GCM and plotted in the same figure with markers. The analytical $S_p$ (dashed curve) matches well with the diagnosed $S_p$ (markers), particularly when $\tau_\infty^s$ is large and convection is shut down. Both RCE and GCM agree that the static stability will significantly increase with $\tau_\infty^s$ when $\tau_\infty^s<4$, followed by a subtle decrease when $\tau_\infty^s>4$. This result does not change qualitatively when choosing other pressure levels, and we show the 900~mb $S_p$ because 900~mb is close enough to the surface to be covered by Hadley cells in all experiments. The transition point $\tau_\infty^s=4=\tau_{LW}$ corresponds to a uniform vertical temperature profile. The subtle decrease of static stability is due to the decrease of the absolute temperature there. This can be seen more clearly if we rewrite the $S_p$ expression as $S_p=\frac{1}{p}(\partial_{z^*}T+\kappa T)$, where $z^*=\ln(p/p_s)$. 

The maximum height of convection or tropopause can be estimated as the level that the dry adiabat profile starting from the ground temperature merges with the radiative equilibrium temperature profile.
\begin{eqnarray}
  T(\tau_{\mathrm{tropp}})\left(\frac{\tau_\infty}{\tau_{\mathrm{tropp}}}\right)^\kappa&=&T_g\nonumber\\
  %\left[1+\chi-(\chi-\chi^{-1})e^{-\tau_{\mathrm{tropp}}/\chi}\right]^{1/4}\left(\frac{\tau_\infty}{\tau_{\mathrm{tropp}}}\right)^\kappa&=&\left[(1+\chi)+(1-\chi)e^{-\tau_\infty/\chi}\right]^{1/4}\nonumber\\
  %\left[1+\chi-(\chi-\chi^{-1})e^{-\tau_{\mathrm{tropp}}/\chi}\right]\tau_{\mathrm{tropp}}^{-4\kappa}&=&\left[(1+\chi)+(1-\chi)e^{-\tau_\infty/\chi}\right]\tau_{\infty}^{-4\kappa}\nonumber\\
  \left[1+\chi-(\chi-\chi^{-1})e^{-\tau_\infty/\chi \frac{p_{\mathrm{tropp}}}{p_s}}\right]\left(\frac{p_{\mathrm{tropp}}}{p_s}\right)^{-4\kappa}&=&\left[(1+\chi)+(1-\chi)e^{-\tau_\infty/\chi}\right],\label{eq:tau-tropp}
  \end{eqnarray}
 where $p_s$ is the surface pressure, $p_{\mathrm{tropp}}$ is the tropopause pressure, and $\kappa=2/7$ for diatomic atmosphere. This estimation is valid only with $\chi>1$, when convection is active.
 
 For an atmosphere transparent to solar radiation, 
  \begin{eqnarray}
    \left(1+\tau^{\chi\rightarrow\infty}_{\mathrm{tropp}}\right)(\tau^{\chi\rightarrow\infty}_{\mathrm{tropp}})^{-4\kappa}&=&(2+\tau_\infty)\tau_{\infty}^{-4\kappa}\nonumber\\
\left(1+\tau_\infty\frac{p^{\chi\rightarrow\infty}_{\mathrm{tropp}}}{p_s}\right)\left(\frac{p^{\chi\rightarrow\infty}_{\mathrm{tropp}}}{p_s}\right)^{-4\kappa}&=&(2+\tau_\infty).    \label{eq:tau-tropp-chi-infty}
\end{eqnarray}
Eq~(\ref{eq:tau-tropp}) and Eq~(\ref{eq:tau-tropp-chi-infty}) can be solved numerically for the tropopause pressure, which can then be used to estimate the tropopause height from $H_{\mathrm{tropp}}^{RCE}=H_{0}\ln{(p_s/p_{\mathrm{tropp}})}$ ($H_0$ is the scale height $RT_0/g$). As a rough estimation for the limits $\chi\rightarrow\infty$, $p^{\chi\rightarrow\infty}_{\mathrm{tropp}}\approx\frac{1}{2}p_s=500$ mb, given that $\kappa\approx 0.25$. In Fig.~\ref{fig:S-Ptropp-RCEvsIsca}(c), numerical solutions for different $\chi$ are plotted in the dashed curve, to be compared with the tropopause heights diagnosed from GCM simulations. The GCM tropopause height is identified by looking for the lowest level with a static stability over $6\times10^{-4}$ K/Pa\footnote{This choice is arbitrary, and thus one cannot reach the conclusion that tropopause height from RCE is fully realistic.}. The match between the analytical solution and the GCM diagnosis indicates the success of the above estimation. Both of them show that convection is suppressed by shortwave absorption, consistent with \cite{Barbaro-Arellano-Krol-et-al-2013:impacts} in the context of the atmospheric boundary layer. One should keep in mind that Eq~(\ref{eq:tau-tropp}) and Eq~(\ref{eq:tau-tropp-chi-infty}) only provide a rough estimate, ignoring the surface temperature adjustment in response to the convection adjustment, as well as other processes such as large-scale dynamics. A more accurate way of doing this is to iterate a 1D RCE model to equilibrium numerically, as done in \cite{Kurokawa-Nakamoto-2012:effects}. But this method is not analytical, and the authors do not explore a full range of $\chi$. We also note that the tropopause under the grey-atmosphere assumption could be lower than is realistic, considering that absorption by real gases often happens in narrow bands, and that latent heating release by condensible species, if present, can rise the tropopause even more by reducing the temperature lapse rate.

\subsection{Circulation response to shortwave absorption}
\label{sec:circulation-response}

\begin{figure*}[htp!]
 \centering
 \includegraphics[width=1\textwidth]{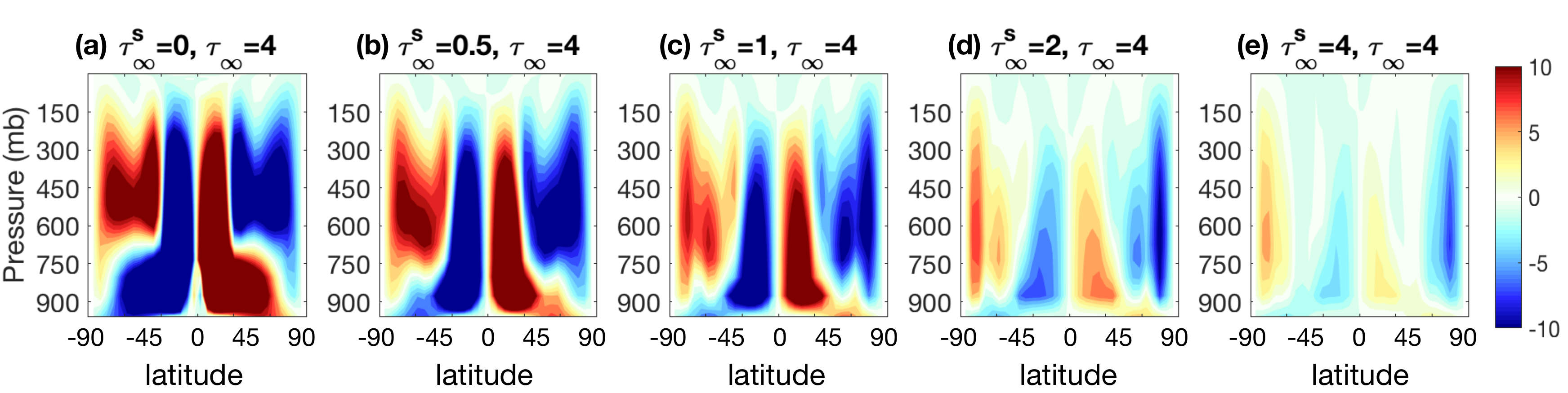}
 \caption{Supression of the meridional circulation at large $\tau_\infty^s$. From left to right, streamfunction $\Psi$ is shown as a function of latitude and pressure for increasing $\tau_\infty^s$, under the ``daily, high-res'' configuration. The relationship between pressure and optical depth is $p=p_s\frac{\tau}{\tau_\infty}$.} 
 \label{fig:Psi-levlat-tausw}
\end{figure*}

In Earth science, meridional streamfunction is used to measure the strength of the zonal mean meridional overturning circulation, and it can be calculated by vertically integrating the meridional wind $V(\theta,p)$, where $\theta,\ p$ are the latitude and pressure.
$$\Psi(\theta,p)=\frac{2\pi a \cos\theta}{g}\int_0^pV(\theta,p)~dp $$
In the above formula, $a$ is the planet's radius and $g$ is the gravitational acceleration rate.
With an enhanced stratification, atmospheric general circulation is suppressed at large $\tau_\infty^s$. This can be clearly seen in Fig.~\ref{fig:Psi-levlat-tausw}, where we show meridional streamfunction $\Psi$ for $0\leq \tau_\infty^s\leq 4$ under the ``daily, high-res'' configurations. 
The general circulation, particularly the Hadley cell, almost vanishes at $\tau_\infty^s>1$. Beyond $\tau_\infty^s=4$, the stratification stops increasing, and thus the circulation remains similar to that in $\tau_\infty^s=4$. We do not show those cases to avoid repetition. While the circulation amplitude changes significantly, the circulation latitudinal extension does not have a clear response.

We note that the meridional circulation with $\tau_\infty^s=0$ in our study looks different from that on Earth. Both the Hadley cell and Ferrel cell are too weak, and the Ferrel cell does not reach to the surface. The former is because we ignore the radiative effect of water vapor by setting the longwave absorptivity to be a constant (i.e., optical depth $\tau$ is linearly proportional to pressure $p$). If we follow \cite{Kaspi-Showman-2015:atmospheric} and set the longwave absorptivity to be larger in low latitudes near the surface to account for a higher water vapor concentration there (not shown), the Hadley circulation is significantly enhanced. The latter issue -- Ferrel cell only exists aloft -- turned out to be related to the dry dynamics used in this study (not shown). With latent heating release, the mid-latitutde baroclinic supercriticality increases \cite[][]{Lapeyre-Held-2004:role}, and as a result, the ageostrophic meridional circulation driven by the baroclinic eddy momentum flux may also be enhanced. In this study, we ignore all these factors to allow us to solve this problem analytically and to compare with idealized GCM simulations.

We next compare the response of the Hadley cell amplitude against two existing theoretical scalings: one proposed by \cite{Held-Hou-1980} and another proposed by \cite{Held-2000:general}, with the aim of predicting the Hadley cell response given the optical depth. Assuming angular momentum conservation along the Hadley upper branch and no further heat transport beyond the Hadley cell, \cite{Held-Hou-1980} predict that Hadley cell extension varies proportionally to
\begin{equation}
  \label{eq:scaling-boundary-HH1980}
  \frac{\sqrt{gH_t\Delta_H}}{\Omega a},
\end{equation}
and the Hadley cell amplitude varies proportionally to
\begin{equation}
  \label{eq:scaling-amplitude-HH1980}
\frac{g^{5/2}H_t^{5/2}\Delta_H^{5/2}}{\tau_{rad}a^2\Omega^3\Delta_V},
\end{equation}
where $H_t$ is the tropical tropopause height, $\Delta_V$ is the vertical potential temperature contrast between the surface and tropopause, $\Delta_H$ is the horizontal potential temperature contrast between the equator and the poles, $\Omega$ is the rotation rate, $a$ is the planet radius, and $\tau_{rad}$ is the radiative time scale. Note that the Held-Hou scaling has been criticized for not capturing the Hadley cell changes in GCM simulations with different configurations \cite[][]{Lu-Vecchi-Reichler-2007:expansion, Wang-Gerber-Polvani-2012:abrupt, Kaspi-Showman-2015:atmospheric}.

% One discrepancy shared by both the Held-Hou and Held-2000 scalings is that the angular momentum is not conserved along the Hadley upper branch in reality. Actually, the eddy momentum transport can play an important role by pumping westerly momentum to mid-latitudes and allowing Hadley circulation to reach farther poleward. It scales well with Hadley amplitude \cite[][]{Walker-Schneider-2006:eddy}, although it has no prediction power without a theory for the mid-latitude eddy activity.

\begin{figure*}[htp!]
 \centering
 \includegraphics[width=0.7\textwidth]{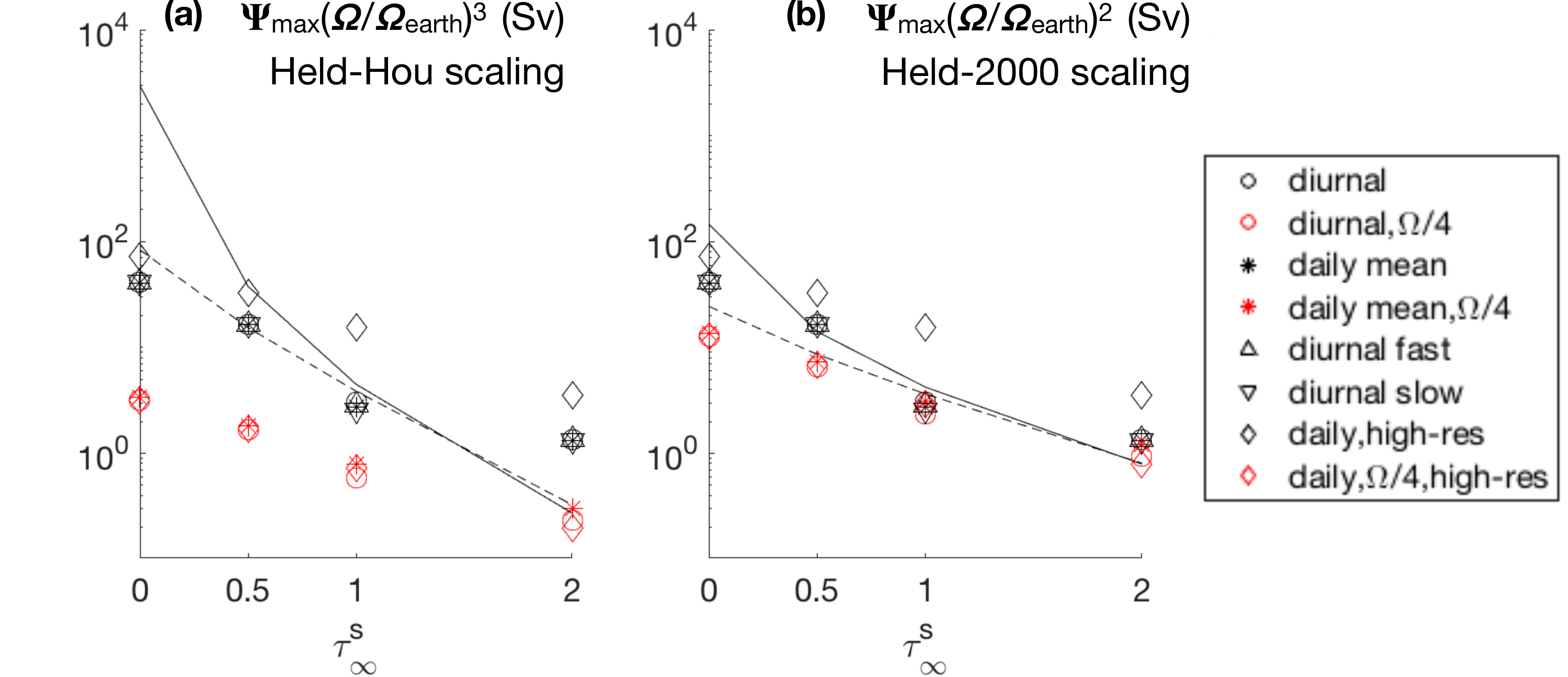}
 \caption{Assessment of the Held-Hou and Held-2000 scalings, by plotting circulation amplitude measured by $\Psi_{max}-\Psi_{max}^{\tau_\infty^s=4}$ as a function of $\tau_\infty^s$ (see text for details), measured in Sverdrups ($1~\mathrm{Sv}=10^9~$kg/s). In panel (a), the circulation amplitude is rescaled by $(\Omega/\Omega_{\mathrm{earth}})^3$, to collapse experiments with different rotation rates following Held-Hou scaling. The scaling for circulation amplitude $\left(H_{\mathrm{tropp}}^{RCE}\right)^{3/2}S_p^{-1}$ is plotted in the black lines. The dashed and solid lines correspond to radiative equilibrium static stability $S_p^{rad}$  and diagnosed static stability, respectively. Different symbols correspond to different experiments as shown in the legend. Panel (b) is the same as panel (a), except that circulation is rescaled by $(\Omega/\Omega_{\mathrm{earth}})^{2}$ as derived from Held-2000 scaling, and the circulation amplitude shown in the black lines is $H_{\mathrm{tropp}}^{RCE}S_p^{-1/2}$. }
 \label{fig:Psimax-tausw}
\end{figure*}

We measure the strength of the meridional circulation in the GCM experiments using the maximum of the meridional streamfunction $\Psi_{max}$ between 40~N/S, minus the $\Psi_{max}$ value at $\tau_\infty^s=4$. $\tau_\infty^s=4$ is chosen as a baseline, because, beyond this point, the tropopause merges with the surface, and the circulation should shut down according to these scalings. The remaining circulation is likely to be driven indirectly by eddy fluxes rather than directly by the thermal contrast, and thus cannot be captured by the scalings. Since the circulation amplitude is predicted to vary with $\Omega^{-3}$, we rescale $\Psi_{max}$ with $(\Omega/\Omega_{\mathrm{earth}})^3$ and show it as a function of shortwave optical depth $\tau_\infty^s$ in Fig.~\ref{fig:Psimax-tausw}(a). After multiplying this $\Omega$ factor, the two fast rotating experiments align with each other, while the $1/4$ rotation experiments does not aligns that well. This may be due to the fact that under the slow rotation limit, the Hadley cell is too wide for the small angle approximation to hold.

Following Held-Hou scaling, the circulation amplitude should vary proportionally with $H_t^{5/2}/\Delta_V \sim H_t^{3/2}/S_p$. We then plug the RCE tropopause height $H_{\mathrm{tropp}}^{RCE}$ and radiative equilibrium static stability $S_p^{rad}$ in and plot the predicted Hadley cell amplitude in the black dashed curve. $H_{\mathrm{tropp}}^{RCE}$ can be solved from Eq.~\ref{eq:tau-tropp}, and $S_p^{rad}$ can be evaluated by substituting the radiative equilibrium temperature in Eq.~\ref{eq:Ta-tau} to the definition $S_p=T\partial_p \ln{\Theta}$, given shortwave and longwave optical depth. The predicted $\Psi_{max}$ vs. $\tau_\infty^s$ slope is reasonable but a bit too large, compared to that diagnosed from the GCMs. When we replace $S_p^{rad}$ with the more realistic $S_p$ diagnosed from GCMs, it begins to diverge signficantly at low $\tau_\infty^s$ (solid black curve). 

With a higher resolution, the meridional circulation gets stronger at Earth's rotation rate (black diamonds in Fig.~\ref{fig:Psimax-tausw}a). This is possibly because eddies are better resolved under high resolution, and these eddies deposit more easterly momentum to the tropics and enhance the poleward motion in the Hadley cell upper branch. As a verification of this argument, we also ran the ``daily mean, $\Omega/4$'' simulations at high resolution. Since eddies play a less important role at slower rotation rate, the difference made by the higher resolution (blue diamonds vs. blue stars) is, as expected, much less significant than in the simulations with Earth rotation rate (black diamonds vs. black stars). In fact, both the Held-Hou scaling and the Held-2000 scaling (to be discussed later) assume that air parcels conserve their angular momentum as going poleward in the Hadley upper branch, which is equivalent to the situation with no zonal resolution at all. Therefore, these scalings are expected to better represent the simulations with a slow rotation or a low resolution. Despite the difference caused by resolution, the meridional circulation suppression by the shortwave absorption remains qualitatively similar to that in the experiments with a lower resolution. 

The other scaling \cite[Held-2000 scaling][]{Held-2000:general} turns out to do a slightly better job in predicting Hadley extension \cite[][]{Lu-Vecchi-Reichler-2007:expansion, Wang-Gerber-Polvani-2012:abrupt}. \cite{Held-2000:general} suggests that the Hadley cell terminates at the latitude that baroclinic instability kicks in, as an air parcel move poleward conserving its angular momentum. This leads to another Hadley terminal latitude scaling:
\begin{equation}
  \label{eq:scaling-boundary-H2000}
  \theta_b\propto\left(\frac{NH_e}{a\Omega^2}\right)^{1/3},
\end{equation}
where $N$ is the Brunt-Vasala frequency, $H_e$ is the extratropical tropopause height. The power becomes $1/2$ if using Phillips two-layer model to get the baroclinic instability condition, instead of the Eady model, but the result does not change qualitatively. \cite{Held-2000:general} does not provide scaling for circulation amplitude. We derive one assuming that the meridional temperature gradient is weak within Hadley circulation. With this assumption, the deviation of the equatorial radiative temperature from radiative equilibrium, $\Delta \Theta(0)=\Theta(0)-\Theta_{rad}(0)$ can be approximated by the radiative equilibrium temperature difference between the equator and the edge of the Hadley cell $\Theta_{rad}(\theta_b)-\Theta_{rad}(0)$. Since the cosine meridional variation of the radiative equilibrium temperature can be approximated by a parabolic function as the insolation has a cosine meridional structure, $\Theta_{rad}(\theta)/\Theta_{rad}(0)=1-\Delta_H\theta^2$, in low latitudes, $\Delta \Theta(0)$ can be approximated by $-\Theta_{rad}(0)\Delta_H\theta_b^2$.
Assuming this deviation $\Delta \Theta(0)$ is supported by upward motion there, we get the scaling for the upward motion,
\begin{eqnarray}
  \omega&=&-\frac{1}{\tau_{rad}\partial_p\Theta}\Delta\Theta(0) \propto \frac{1}{\tau_{rad} S_p}\Delta\Theta(0). \label{eq:omega-scaling-Held2000}
\end{eqnarray}
Then the meridional streamfunction amplitude scales as
\begin{equation}
  \Psi_{max}\propto -\omega a\theta_b=-\frac{a}{\tau_{rad} S_p}\Delta\Theta(0)\theta_b=\frac{a \Theta_{rad}(0)\Delta_H}{\tau_{rad} S_p}\theta_b^3\nonumber
\end{equation}
Substituting the Held-2000 scaling for the Hadley boundary latitude $\theta_b$, we get
\begin{equation}
  \Psi_{max} \propto \frac{a \Theta_{rad}(0)\Delta_H}{\tau_{rad} S_p}\frac{NH_e}{a\Omega^2}\propto \frac{\Theta_{rad}(0)\Delta_HH_e}{\tau_{rad} S_p^{1/2}\Omega^2}\label{eq:Psimax-scaling-Held2000}
\end{equation}

The usage of extratropical tropopause height $H_e$ can be tricky, because the same baroclinic instability criteria can be used to solve for $H_e$ inversely, which would make the predictive power of the formula an illusion. To retain predictive power, we here approximate $H_e$ again using the RCE tropopause height $H_{\mathrm{tropp}}^{rad}$\footnote{The RCE tropopause heights from Eq~(\ref{eq:tau-tropp}) and Eq~(\ref{eq:tau-tropp-chi-infty}) are, by definition, more closely linked to the tropopause height in the tropics where convection is active.}.

Following the Held-2000 scaling, we rescale $\Psi_{max}$ with $(\Omega/\Omega_{\mathrm{earth}})^2$, and plot it against $\tau_\infty^s$ in Fig.~\ref{fig:Psimax-tausw}(b). This rescaling does a better job in collapsing the $\Psi_{max}$ curve from experiments with different rotation rates than the Held-Hou scaling. The circulation amplitude variation with $\tau_\infty^s$ is well predicted by the scaling $H_{\mathrm{tropp}}^{rad}S_p^{-1/2}$, especially when using the diagnosed static stability (solid curve). Again, high resolution does not significantly affect the suppression of the meridional circulation by the atmospheric shortwave absorption, although the circulation is enhanced in general.

To conclude, both the Held-Hou and Held-2000 scalings do a reasonable job in predicting the response of Hadley circulation amplitude to changes in rotation rates and shortwave optical depth. The diurnal cycle (movement of substellar point) does not have significant impacts on the tropospheric general circulation. The circulation change may have observable consequences, such as tropical cloud coverage and superrotation (discussed in section~\ref{sec:superrotation}), as suggested in \cite{Kang-2019:regime}.

We have also checked the response of the Hadley cell width against the two scalings (Eq.~\ref{eq:scaling-boundary-HH1980}, Eq.~\ref{eq:scaling-boundary-H2000}), but, only in the slow rotation experiments (diurnal $\Omega/4$, daily-mean $\Omega/4$ and daily $\Omega/4$ high-res), the Hadley cell shrinks with $\tau_\infty^s$ as predicted. Under the Earth rotation rate, the Hadley cell, instead, tends to expand with $\tau_\infty^s$, and the amount of the expansion varies significantly with the horizontal resolution, indicating that eddies play a key role in determining the Hadley boundary. It could be that the Hadley edge is primarily determined by where the Ferrel cell (which is driven by baroclinic eddies) becomes strong enough to overwhelm the Hadley cell. This physical picture is consistent with the original Held-2000 scaling, and thus one would expect the Held-2000 scaling should be able to predict the Hadley cell boundary as shown in \cite{Lu-Vecchi-Reichler-2007:expansion}. The inconsistency arising in our work may be because we replace the extratropical tropopause height $H_e$ with the RCE tropical tropopause height $H_{\mathrm{tropp}}^{rad}$ in Eq.~(\ref{eq:scaling-boundary-H2000}) to retain prediction power. Since this is not our main focus in this work, we do not discuss or show the results in further detail.

The failure of the two scalings in predicting the Hadley width does not necessarily exclude their capability in capturing the response of Hadley cell amplitude to shortwave opacity, because, in both scalings, vertical motions will be prohibited by strong stratification. Again, with the simulations in our study, we cannot, and do not intend to, claim either scaling to be better than the other, instead, we attempt to demonstrate that both scalings captures the first order response of Hadley cell amplitude to the changes of shortwave optical depth.

\subsection{Superrotation}
\label{sec:superrotation}

\begin{figure*}[htp!]
 \centering
 \includegraphics[width=0.9\textwidth]{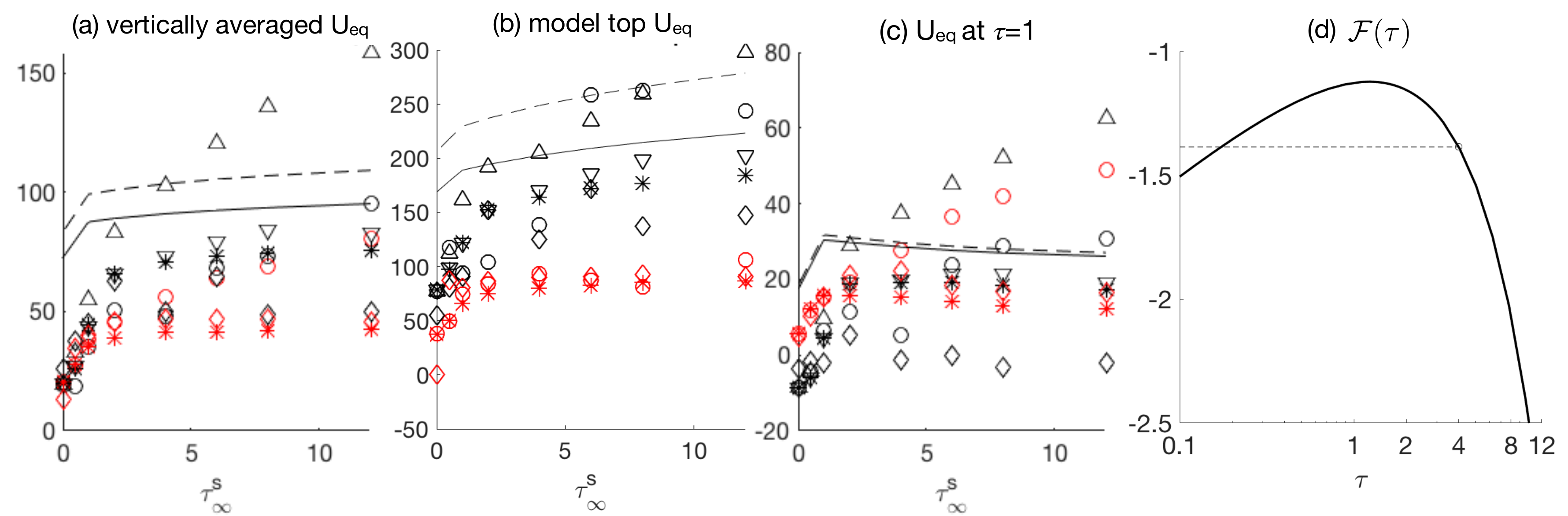}
 \caption{Panel (a-c) shows the superrotation response to the atmospheric shortwave optical depth $\tau_\infty^s$, measured by (a) equatorial zonal wind vertically averaged in $z$, (b) equatorial zonal wind at the model top, (c) equatorial zonal wind at $\tau=1$. The wind speed is rescaled by $\Omega/\Omega_{\mathrm{earth}}$. For comparison, the gradient wind speeds $U_{rad}$ (Eq.~\ref{eq:Urad}) are shown in the solid curves, and the thermal wind speeds (drop the curvature term in Eq.~\ref{eq:Urad}) are shown in the dashed curves. The symbol definitions are the same as Fig.~\ref{fig:Psimax-tausw}. Panel (d) shows the function $\mathcal{F}(\tau)$ defined in Eq.~(\ref{eq:Urad-chi-transition}). For details, please refer to the main text.}
 \label{fig:superrotation-tau-diurnal-cycle}
\end{figure*}

Wind speeds in the upper atmospheres of rocky exoplanets may potentially be detected through high-resolution Doppler spectrometry \cite[][]{Snellen-De-De-et-al-2010:orbital, Showman-Fortney-Lewis-et-al-2012:doppler}. In this section, we discuss how the equatorial super-rotation in the upper atmosphere is affected by the shortwave absorptivity and diurnal cycle.
Fig.~\ref{fig:superrotation-tau-diurnal-cycle} summarizes the results. The vertically averaged equatorial zonal wind is diagnosed from GCM simulations under various configurations, rescaled by $\Omega/\Omega_{\mathrm{earth}}$, and plotted as a function of shortwave optical depth $\tau_\infty^s$. In all experiments, we find superrotation above the equator, as noticed by many studies including the pioneering work by \cite{Young-Pollack-Young-1977:three}.

%Since there is no grid point right above the equator, we use the wind speed at the two grid points that are closest to the equator at T21 resolution, which is 2.8N/S. Since 2.8N/S is slightly off the equator, it may or may not be a good indicator for the equatorial zonal wind, and this is clearly an unavoidable caveat of using spectral model. However, on the other hand, we may be able to use the gradient wind balance of thermal wind balance to understand the wind speed there. The main difference between the two balances is that the gradient wind balance has an extra curvature term $U^2\tan\theta/a$ in addition to the geostrophic terms. This extra term evaluated at 2.8N/S would be 5 times smaller than the Coriolis term $2\Omega \sin\theta U$, even when substituting a wind speed as large as 200~m/s. We therefore stick to the thermal wind balance in most analysis here.

The formation of equatorial superrotation requires extra momentum convergence by eddies \cite[][]{Read-Lebonnois-2018:superrotation}, according to Hide's theorem -- angular momentum extrema cannot exist in the atmosphere interior without eddy fluxes. These eddies can be generated through barotropic instability, according to the classical Girasch-Rossow-Williams mechanism \cite[][]{Gierasch-1975:meridional, Rossow-Williams-Rossow-1979:large}, or by the zonally asymmetry of the heating above the equator, according to the ``moving flame'' mechanism \cite[][]{Schubert-Whitehead-1969:moving, Schubert-1983:general}, by equatorial waves \cite[][]{Showman-Polvani-2010:matsuno, Showman-Polvani-2011:equatorial}, or by thermal tides \cite[][]{Pechmann-Ingersoll-Pechmann-1984:thermal}.

Although it does not elucidate the forcing mechanism directly, gradient wind balance can be used to predict the strength of equatorial superrotation in the GCM. We write the full balance as 
%\footnote{The gradient wind balance helps us predict superrotation from the aspect of keeping balance, but cannot explain how it forms.}.
\begin{eqnarray}
  \label{eq:gradient-wind-balance}
  f_{0}\sin\theta U(\theta,p)+\frac{\tan\theta}{ a}U(\theta,p)^2&=&-g/a\partial_\theta Z(\theta,p).
\end{eqnarray}
The above equation is written in $p$ coordinate. $f_0=2\Omega$ is the Coriolis coefficient at the equator. $Z$ is the geopotential height, equivalent to the pressure in $z$ coordinate. It can be solved by vertically integrating the temperature field from some bottom pressure $p_b$,
\begin{eqnarray}
  \label{eq:integrate-Z}
  Z(\theta,p)=Z(\theta, p_b)+\int_{p}^{p_b} RT(\theta,p')~\frac{dp'}{p'}.
\end{eqnarray}
For the theory to have predictive power, the meridional gradient of both the temperature profile $T$ and the geopotential height $Z$ at the bottom $p_b$ have to be known beforehand. An obvious choice of temperature profile is the radiative-convective equilibrium (RCE) temperature calculated in Section~\ref{sec:RCE-analytical}. For $p_b$, the surface pressure is an obvious choice. However, it is not a good one, even though we know the geopotential height there must be zero. First, the surface pressure is not uniform: strong gradients will develop with the Hadley cell to overcome the surface friction. Second, surface friction can break gradient wind balance in the planetary boundary layer. Third, in the troposphere, where the general circulation is strong, the temperature profile can be far from RCE.
Hence a better choice of $p_b$ is the minimum of the boundary layer pressure and the tropopause pressure,
\begin{eqnarray}
  \label{eq:pb}
  p_b=\min[p_{bnd},p_{tropp}],
\end{eqnarray}
where $p_{bnd}$ is 700~mb and $p_{tropp}$ can be calculated from Eq.~(\ref{eq:tau-tropp}).
This choice has none of the above flaws. It is an isobar (a surface with constant pressure), as its expression (Eq.~\ref{eq:tau-tropp}) only depends on the optical depth, which does not vary with latitude in our setup. Above the boundary layer, surface friction is zero and gradient wind balance is well satisfied. Above the tropopause, atmospheric dynamic time scale becomes longer than radiative time scale. As a result, the temperature profile is very close to the RCE profile, or equivalently, the radiative equilibrium temperature profile (Eq.~\ref{eq:Ta-tau}), in most places (not shown)\footnote{The polar regions are an exception, likely because Rossby waves transmitted from below transport a significant amount of heat poleward there \cite[][]{Vallis-2006:atmospheric}.}.

We set the meridional gradient of the geopotential height to zero everywhere below $p_b$. This is equivalent to setting $U$ to zero below $p_b$ in the gradient wind framework (Eq.~\ref{eq:gradient-wind-balance}), which can be justified as following: within the boundary layer, the surface friction can efficiently damp $U$ to zero, and these zero angular momentum air parcels can quickly fill the whole troposphere at the equator due to transport by the Hadley cell, before their momentum can be modified by eddy momentum transport. The vanishing of geopotential height gradient below the tropopause is consistent with the weak pressure gradient in the tropical upper troposphere proposed by \citet{Romps-2012:weak}. 

With these assumptions, we can analytically solve for a zonal wind profile $U_{rad}$, given the radiative equilibrium temperature profile $T_{rad}$ (Eq.~\ref{eq:Ta-tau}) and the tropopause pressure (Eq.~\ref{eq:tau-tropp}),
\begin{eqnarray}
  &~&U_{rad}+\frac{1}{2\Omega a\cos\theta}U_{rad}^2=-\frac{R}{af_{0}\sin\theta}\partial_\theta\int_{p}^{p_b} T_{rad}~\frac{dp}{p}\nonumber\\
        % &=&-\frac{R}{af_{0}\sin\theta}\partial_\theta T_{\mathrm{skin}}(\theta) \int_{\tau}^{\tau_b} \left[1+\chi-(\chi-\chi^{-1})e^{-\tau/\chi}\right]^{1/4}~\frac{d\tau}{\tau}\nonumber\\
                 &=&-\frac{R}{af_{0}\sin\theta}\partial_\theta \left(\frac{S_0\cos\theta}{2\sigma}\right)^{1/4} \int_{\tau}^{\tau_b} \left[1+\chi-(\chi-\chi^{-1})e^{-\tau/\chi}\right]^{1/4} ~\frac{d\tau}{\tau}\nonumber\\
  &=&\frac{R}{4af_{0}} \left(\frac{S_0}{2\sigma\cos^3\theta}\right)^{1/4} \int_{\tau}^{\tau_b} \left[1+\chi-(\chi-\chi^{-1})e^{-\tau/\chi}\right]^{1/4} ~\frac{d\tau}{\tau}.\label{eq:Urad}
\end{eqnarray}
Note that, after substituting $T_{rad}$, there is no singular point at the equator even though the Coriolis coefficient $f_0\sin\theta$ vanishes there. There are two solutions, one positive and one negative, for the above quadratic equation. The negative solution corresponds to a dominant balance between the pressure gradient force and the curvature term, and is therefore unphysical in a system that is close to thermal wind balance.

In reality, the meridional temperature gradient in the tropical stratosphere could be slightly homogenized, the atmospheric dynamics may play a significant role beyond $p_b$ until a much higher level\footnote{Fig.~\ref{fig:Psi-levlat-tausw} shows that the Hadley cell can extend to 150~mb, but the predicted tropopause is below 622~mb for all cases.}, and the baroclinic eddies in the mid-latitudes can transport westerly momentum poleward. All these lead to deceleration of equatorial wind. Therefore, the $U_{rad}$ at the equator sets the upper bound of the equatorial superrotation, unless other external forcings (e.g., diurnal cycle) significantly redistribute momentum or temperature in the stratosphere.

\paragraph{The equatorial superrotation in cases without a diurnal cycle,}

$U_{rad}$ at the equator is shown by solid curves in Fig.~\ref{fig:superrotation-tau-diurnal-cycle}(a,b,c) as a function of the shortwave optical depth $\tau_\infty^s$ for three different levels: vertically averaged, model top and unit longwave optical depth $\tau=1$. Overlaid on these plots are the equatorial wind diagnosed from the GCM simulations. Despite that $U_{rad}$ tends to overestimate the equatorial superrotation as expected, it captures the superrotation's dependence on $\tau_\infty^s$ in the cases without a diurnal cycle reasonably well (marked by stars or diamonds). The superrotation wind speed is normalized by a factor of $\Omega/\Omega_{\mathrm{earth}}$ to account for the different Coriolis coefficients at different rotation rates. Hereafter, we refer to this speed the normalized superrotation speed.

In the limit of weak shortwave absorption ($\tau_\infty^s\rightarrow 0$), both the analytical prediction and GCM simulations show a dip at all three levels. This is because the higher tropopause height in those cases causes a larger portion of the atmosphere to have vanishing $U_{rad}$ and $\partial_\theta Z$.
In the limit of strong shortwave absorption (large $\tau_\infty^s$), the dependence of superrotation on $\tau_\infty^s$ varies with the levels. At higher altitudes (e.g., the model top level shown in panel b), the equatorial superrotation strengthens with $\tau_\infty^s$, while at lower altitudes (e.g., the unit longwave optical depth level shown in panel c), it weakens with $\tau_\infty^s$ instead. To understand this, we Taylor expand the right hand side of Eq.~(\ref{eq:Urad}) around $\tilde{\chi}\equiv \chi^{-1}\rightarrow 0$,
\begin{eqnarray}
  &~&U_{rad}+\frac{1}{2\Omega a\cos\theta}U_{rad}^2\nonumber\\
  &=& \frac{R}{4af_{0}} \left(\frac{S_0}{2\sigma\cos^3\theta}\right)^{1/4} \left[\int_{\tau}^{\tau_b}\left(1+\tilde{\chi}^{-1}+(\tilde{\chi}-\tilde{\chi}^{-1})(1-\tau\tilde{\chi}+\frac{1}{2}\tau^2\tilde{\chi}^2))\right)^{1/4}~\frac{d\tau}{\tau}\right]  \nonumber\\
  &=& \frac{R}{4af_{0}} \left(\frac{S_0}{2\sigma\cos^3\theta}\right)^{1/4} \left[\int_{\tau}^{\tau_b}\left(1+\tau+\tilde{\chi}-\frac{1}{2}\tau^2\tilde{\chi}\right)^{1/4}~\frac{d\tau}{\tau}\right]  \nonumber\\
  &=& \frac{R}{4af_{0}} \left(\frac{S_0}{2\sigma\cos^3\theta}\right)^{1/4} \left[\int_{\tau}^{\tau_b}\frac{(1+\tau)^{1/4}}{\tau}~d\tau +\tilde{\chi}\int_{\tau}^{\tau_b}\frac{1}{8\tau(1+\tau)^{3/4}}(2-\tau^2)~d\tau\right]  \label{eq:Urad-taylor}
\end{eqnarray}
Whether $U_{rad}$ increases or decreases with $\tilde{\chi}=\tau_\infty^s/\tau_\infty$ depends on the sign of the second integral. This integrated function has a singularity at $\tau\rightarrow 0$, meaning that, as long as $\tau$ (or equivalently, $p$) is small enough, the integral is always positive, and it approaches infinity with $\log$ dependence as the observed level $p\rightarrow 0$. However, in reality, we can only observe the wind speed at a finite pressure level ($p=0$ is infinitely far away), and the integral can be negative with a large enough $\tau_b$. The critical $\tau_b$ can be calculated given $\tau$ by setting the second integral in Eq.~\ref{eq:Urad-taylor} to zero,
\begin{eqnarray} 
&~&\int_{\tau}^{\tau_b}\frac{1}{\tau(1+\tau)^{3/4}}(2-\tau^2)~d\tau=0\nonumber\\
&~&\int_{\tau}^{\tau_b}\frac{1}{\tau}(2-\tau^2)~d(1+\tau)^{1/4}=0\nonumber\\
&~&\int_{(1+\tau)^{1/4}}^{(1+\tau_b)^{1/4}}\frac{2}{x^4-1}-(x^4-1)~dx=0, ~~\mathrm{where}~~x\equiv(1+\tau)^{1/4} \nonumber\\
&~&\left\{\frac{1}{4}\ln\left[\frac{(1+\tau)^{1/4}-1}{(1+\tau)^{1/4}+1}\right]-\arctan\left[(1+\tau)^{1/4}\right]-\frac{1}{10}(\tau+1)^{1/4}(\tau-4)\right\}_{\tau}^{\tau_b}=0 \label{eq:Urad-chi-transition}
\end{eqnarray}

We denote the function within the square bracket as $\mathcal{F}(\tau)$ and show it in Fig.~\ref{fig:superrotation-tau-diurnal-cycle}d. For any given $\tau_b$ (here $\tau_b\lesssim 4$), one can find corresponding $\mathcal{F}(\tau_b)$ on the curve, draw a horizontal line away from the curve. The other intersection with the $\mathcal{F}(\tau_b)$ curve will be the critical $\tau_{crit}$. For $\tau_b=4$, $\tau_{crit}=0.173$, which corresponds to $p_{crit}=43.3$~mb. The model top (6~Pa) is above the $p_{crit}$, and as expected, the equatorial superrotation in the ``daily mean'' experiments (denoted by stars) enhances with $\tau_\infty^s$ in general (Fig.~\ref{fig:superrotation-tau-diurnal-cycle}b). The $\tau=1$ level (250~mb) is below the $p_{crit}$, and also as expected, the equatorial superrotation in the ``daily mean'' (denoted by stars) experiments weakens with $\tau_\infty^s$ except the dip at small $\tau_\infty^s$ (Fig.~\ref{fig:superrotation-tau-diurnal-cycle}c).

Experiments at a faster rotation rate (black stars and diamonds) tend to have a faster normalized superrotation speed compared to those with a smaller rotation rate (red stars and diamonds). This is because the general circulation (both in the eddy form and in the mean flow form) is suppressed for a larger Coriolis coefficient, leading to a greater meridional temperature gradient as well as weaker baroclinic eddies \cite[][]{Kaspi-Showman-2015:atmospheric}. A greater meridional temperature gradient requires faster westerly wind to balance, and weaker eddies transport less westerly momentum toward the mid-latitude baroclinic zone, both enhancing the super-rotation above the equator. Some studies \cite[][]{Laraia-Schneider-2015:superrotation, Polichtchouk-Cho-2016:equatorial} have shown that superrotation is suppressed when the meridional temperature gradient is strong, but that is mainly because the mid-latitude baroclinic eddies are enhanced by the stronger meridional temperature gradient.

Now that we understand how equatorial superrotation changes with the shortwave absorption in atmosphere using gradient wind balance, we consider the even simpler thermal wind balance, where the curvature term (second term in Eq.~\ref{eq:Urad}) is ignored. The equatorial superrotation predicted by thermal wind is shown by dashed curves in Fig.~\ref{fig:superrotation-tau-diurnal-cycle}(a,b,c). Its main features are the same as the prediction based on gradient wind balance, except that it overestimates the equatorial superrotation in GCMs even more, indicating that the thermal wind balance alone may be a good enough approximation. 

Equatorial superrotation appears in all experiments here (Fig.~\ref{fig:superrotation-tau-diurnal-cycle}). However, this superrotation, especially that in the ``daily mean'' cases, could be an artifact of the idealized physics scheme used here. For example, when we set the surface albedo to 0.25 and set the optical depth to be thicker at low latitudes in order to mimic the radiative effect of water vapor as done in \cite{Kaspi-Showman-2015:atmospheric}, the radiative equilibrium temperature in the upper atmosphere is lower than the surrounding area, and thus the equatorial wind at the model top turns easterly.

Insufficient spatial resolution has been found to cause artificial superrotation above the equator \cite[][]{Polichtchouk-Cho-2016:equatorial}. We repeat the ``daily mean'' experiments under T42 resolution\footnote{\cite{Polichtchouk-Cho-2016:equatorial} indicates that T170 resolution may be required for the superrotation to fully converge under a weakened meridional temperature gradient, while T42 resolution may be enough under a regular meridional temperature gradient.}, and the results are shown in the same plot using diamonds. At a quarter Earth rotation rate, doubling the resolution does not affect the result significantly. At the Earth rotation rate, doubling the resolution does weakens the equatorial wind speed, but the conclusions does not change: superrotation increases with $\tau_\infty^s$ in the upper atmosphere, and weakens with $\tau_\infty^s$ in lower atmosphere.

\paragraph{The diurnal cycle effects}

 \begin{figure*}[htp!]
 \centering
 \includegraphics[width=0.8\textwidth]{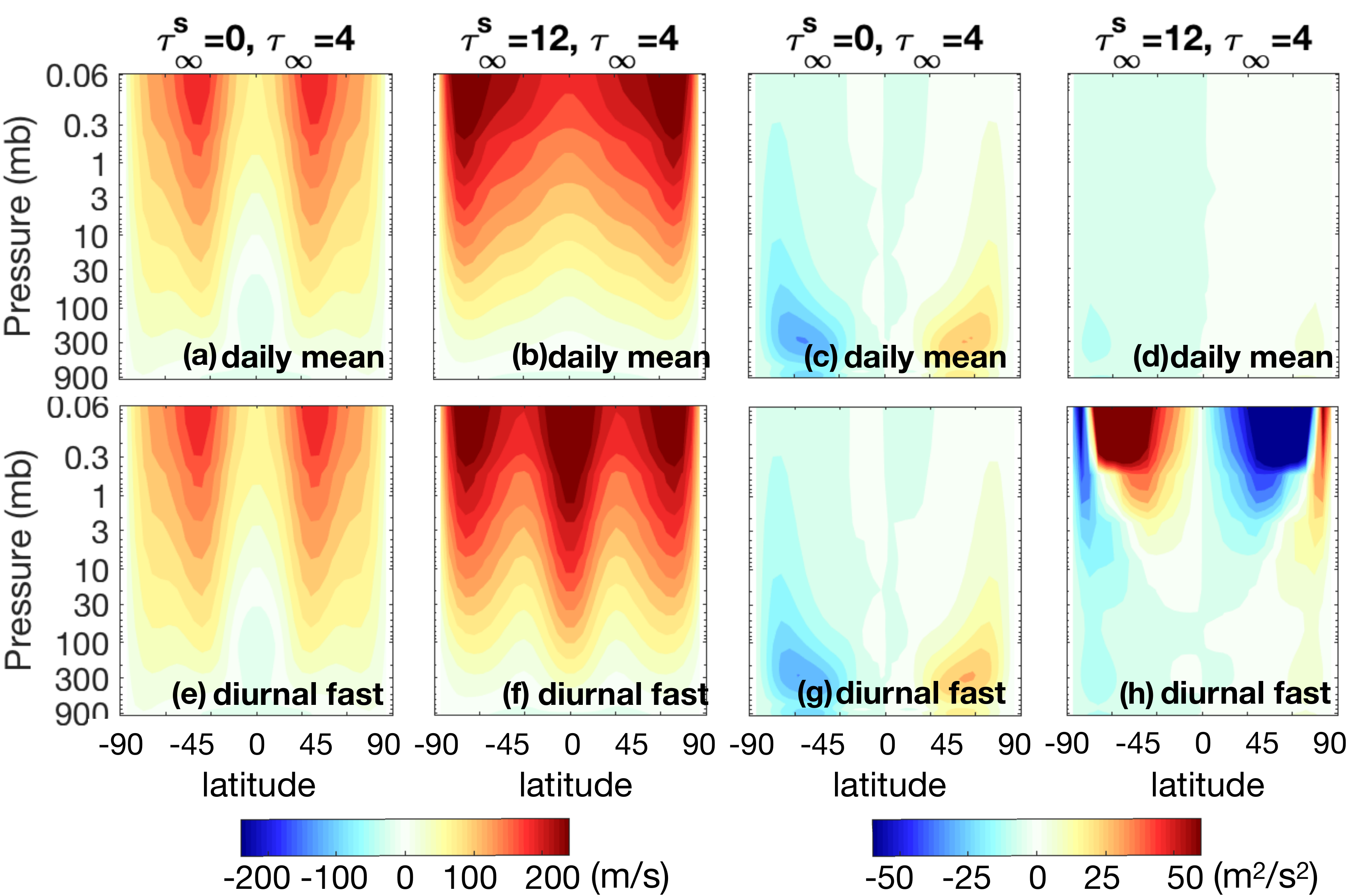}
 \caption{Zonal mean zonal wind (left four panels) and zonal mean meridional eddy momentum transport $\overline{u'v'}$ (right four panels) for four experiments. Shown are for (a,c) ``daily mean'' experiment with $\tau_\infty^s=0$,  (b,d) ``daily mean'' experiment with $\tau_\infty^s=12$, (e,g) ``diurnal fast'' experiment with $\tau_\infty^s=0$, (f,h) ``diurnal fast'' experiment with $\tau_\infty^s=12$. The relationship between pressure and optical depth is $p=p_s\frac{\tau}{\tau_\infty}$.}.
 \label{fig:superrotation-diurnal-cycle-mechanism}
\end{figure*}

$U_{rad}$ should be the upper limit of the equatorial zonal wind for the ``daily mean'' experiments (denoted by stars and diamonds). However, with diurnal cycle (sun moving westward), the ``moving flame'' effect \cite[][]{Schubert-Whitehead-1969:moving, Schubert-1983:general}, the equatorial Rossby waves  \cite[][]{Showman-Polvani-2010:matsuno, Showman-Polvani-2011:equatorial}, and the thermal tide \cite[][]{Pechmann-Ingersoll-Pechmann-1984:thermal} can converge westerly momentum equatorward, accelerating the equatorial zonal wind beyond $U_{rad}$. We would expect these effects to be stronger with a larger $\tau_\infty^s$, as the shortwave radiation start to be directly absorbed by the upper atmosphere. An alternative way to put this is that when the radiative timescale gets shorter than the diurnal cycle at high $\tau_\infty^s$, the temperature will not be homogeneous zonally; this inhomogeneity can excite equatorial waves converging westerly momentum equatorward.

As shown in Fig.~\ref{fig:superrotation-tau-diurnal-cycle}(a), most ``diurnal'' experiments have greater superrotation than the corresponding ``daily mean'' experiments. We show the zonal mean zonal wind distribution for two extreme examples in the left four panels of Figs.~\ref{fig:superrotation-diurnal-cycle-mechanism}. With $\tau_\infty^s=0$, even a fast diurnal cycle does not make any difference (panels a,e), while with $\tau_\infty^s=12$, the fast diurnal cycle leads to almost 150 m/s higher upper air speeds above the equator (panels b,f). The superrotation strengthening is seen in all shortwave absorbing cases and it gets stronger with the shortwave absorptivity. This is because only with a non-zero shortwave absorptivity, the movement of the sun induced by the diurnal cycle can be directly "felt" by the upper atmosphere, which in turn triggers the Gill response \cite[][]{Gill-1980:some} and converges westerly momentum equatorward. This Gill response has been applied to explain the superrotation on tidally-locked planets \cite[][]{Showman-Polvani-2010:matsuno, Showman-Polvani-2011:equatorial, Arnold-Tziperman-Farrell-2012:abrupt}. Shown in right four panels of Fig.~\ref{fig:superrotation-diurnal-cycle-mechanism} are the corresponding eddy meridional momentum transport $\overline{u'v'}$. Only in Fig.~\ref{fig:superrotation-diurnal-cycle-mechanism}h, a significant equatorward momentum convergence is evident, and this is the main driver of the eastward acceleration seen in Fig.~\ref{fig:superrotation-diurnal-cycle-mechanism}f (we also checked that $\overline{u'\omega'},\ \overline{v'T'}$ were far weaker; results not shown). The upper atmosphere in the other three cases cannot directly "feel" the movement of the sun induced by the diurnal cycle, in absent of either shortwave absorptivity or diurnal cycle, and thus don't have eddy momentum convergence close to the model top (panels c,d,g in Fig.\ref{fig:superrotation-diurnal-cycle-mechanism}).

\section{Conclusions}
\label{sec:conclusions}

\begin{figure*}[htp!]
 \centering
\includegraphics[width=0.65\textwidth]{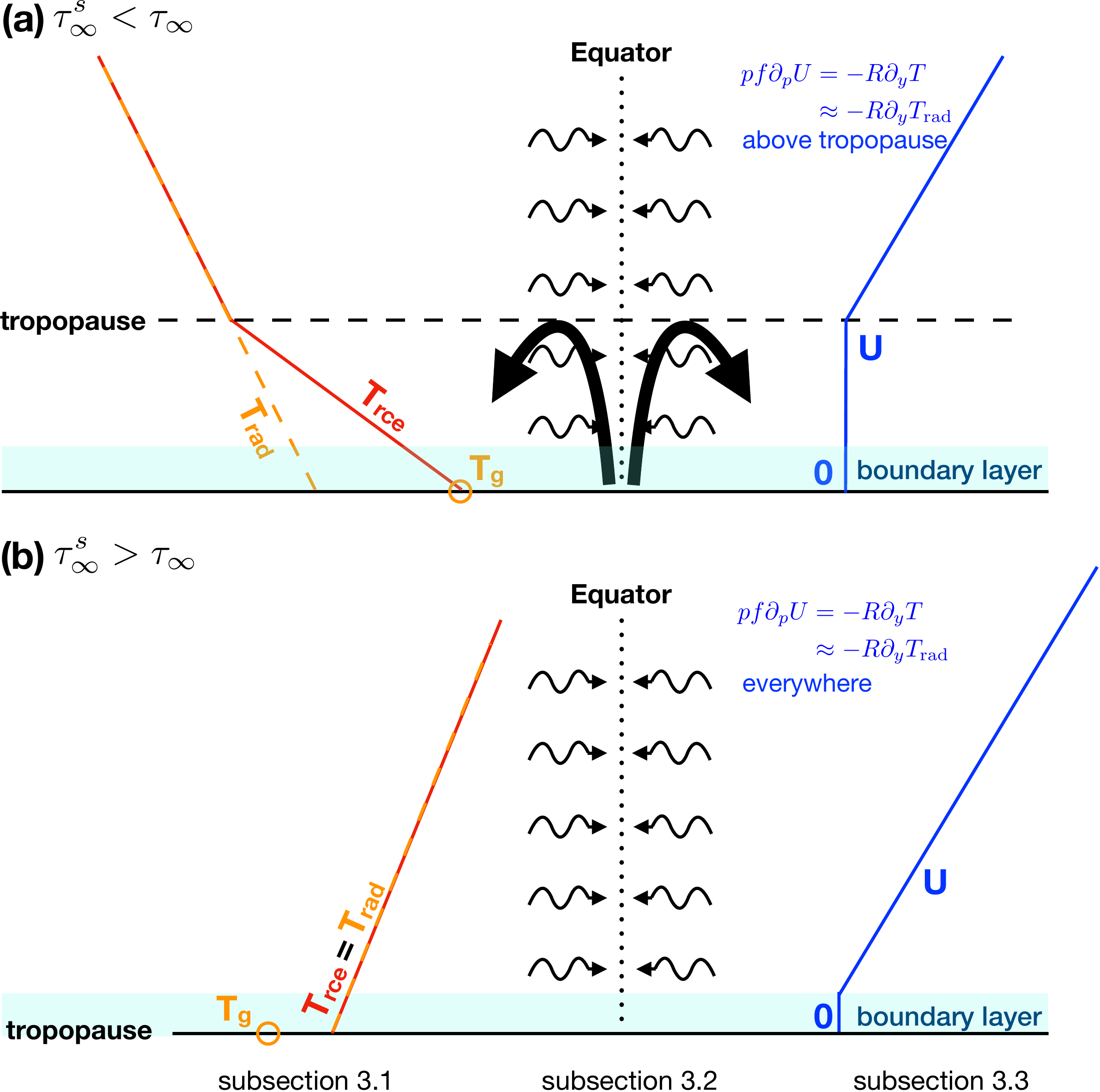}
 \caption{Schematics sketching the physics considered in this study. From left to right, we sketch the key processes that affect radiative transfer, meridional overturning general circulation and the equatorial superrotation, which will be discussed in subsection~\ref{sec:RCE-analytical}, \ref{sec:circulation-response} and \ref{sec:superrotation}. The case with relatively smaller shortwave absorption ($\chi\equiv\tau_\infty/\tau_\infty^s>1$) is shown in panel (a), and the opposite case ($\chi<1$) is shown in panel (b). Details are in the main text.} 
 \label{fig:overview}
\end{figure*}

%% what we have done and main results summary
In this study, we investigated how the circulation and equatorial supperrotation respond to atmospheric shortwave absorption on a terrestrial-type planet analytically, and compared the results with 3D GCM simulations. Our focus is mostly put on dry atmospheres, without condensible components. Fig.~\ref{fig:overview} sketches the key findings in our work. The relative magnitude of the longwave optical depth $\tau_\infty$ and shortwave optical depth $\tau_\infty^s$ controls whether temperature decreases or increases with altitude and whether the surface temperature is warmer or colder than surface air temperature, leading to two separate regimes, as shown in the two panels of Fig.~\ref{fig:overview}.

We first solve the temperature profile analytically and make a prediction of the tropopause height based on the temperature profile; both well match the GCM simulations. With relatively weak shortwave absorption (panel a), solar radiation reaches the surface without too much loss, and the radiative equilibrium temperature profile is unstable for convection in the lower troposphere. With strong atmospheric shortwave absorption (panel b), incoming radiation is absorbed before reaching the surface, leading to a temperature profile that is stable for convection. As $\tau_\infty^s\rightarrow \infty$, the surface temperature even approaches the skin temperature. As a result, the habitable zone may become shifted inward significantly. 

We then use the analytical temperature profile and the tropopause height to predict the Hadley cell amplitude using the Held-Hou scaling \citep{Held-Hou-1980} and the Held-2000 scaling \citep{Held-2000:general}. Both correctly predict the suppression of Hadley cell as the atmospheric stratification strengthens with the shortwave optical depth. When $\tau_\infty^s$ becomes greater than $\tau_\infty$, the convection and Hadley cell are completely suppressed, as depicted in the middle part of Fig.~\ref{fig:overview}(b).

We finally try to predict the equatorial superrotation by solving for the gradient wind in balance with the radiative equilibrium temperature. The formation of equatorial superrotation requires eddies. Without a diurnal cycle or moist effects, the equatorial eddy activity is in general weak (see Fig.~\ref{fig:superrotation-diurnal-cycle-mechanism}c,d), and thus can be dominated by the surface friction and the vertical advection of the Hadley cell in places where they get strong. Within the boundary layer, the surface friction efficiently damps the wind speed to zero; the Hadley cell then quickly fills the entire equatorial troposphere with this zero speed air (see Fig.~\ref{fig:superrotation-diurnal-cycle-mechanism}a,b). The vanishing of zonal wind there also indicates a weak pressure gradient as suggested by \citet{Romps-2012:weak}. Only beyond the tropopause and the boundary layer, the temperature profile remains close to the radiative equilibrium, and the zonal wind starts to accelerate with altitude as required by the gradient wind balance or thermal wind balance (see the right part of Fig.~\ref{fig:overview}). The diurnal cycle can further accelerate the equatorial superrotation by exciting equatorial Rossby waves.

% observables
A weaker Hadley circulation may reduce cloud formation above the equator in the Inter-Tropical Convergence Zone (ITCZ), which is potentially observable. Another observable consequence of shortwave absorption is the strengthening of superrotation in the upper atmosphere above the equator with $\tau_\infty^s$. On top of this, the diurnal cycle can further strengthen the superrotation by exciting equatorial waves, particularly when the shortwave absorption is strong. 

%% caveat
Although a grey radiation GCM without condensation is easier to compare with analytical solutions, it may lead to an underestimation of the tropopause height and hence the circulation, because otherwise the longwave absorption constrained in narrow bands can cool the stratosphere below the skin temperature, and latent heat release can significantly reduce the temperature lapse rate. Nevertheless, the conclusion that the general circulation can be suppressed in a shortwave-absorbing atmosphere is expected to hold. This circulation change is not directly observable, but it may be observable indirectly via a reduction in cloud formation above the equator when there is condensible component in the atmosphere. However, moist processes, including cloud formation, are ignored in this study, so future work is needed to investigate these effects. Equatorial superrotation appears in all simulations in this study. On one hand, this indicates that the equatorial superrotation may be quite common. On the other hand, our highly idealized model setup, in particular, the missing of water vapor and obliquity effects, may allow superrotation to form more easily. 

% a happy ending with perspective
The lesson we learn from this highly idealized study may be used to understand the atmospheric dynamics on planets with strong shortwave atmospheric absorption. In the solar system, such examples include Venus, Titan and Mars during dust storm \citep{Read-Barstow-Charnay-et-al-2016:global}. With a significant amount of solar radiation absorbed by the atmosphere, we may expect a very stable layer formed near the surface to suppress the general circulation there. Meanwhile, we expect the shortwave radiation absorbed in the top of the atmosphere to drive a stronger equatorial superrotation, particularly in appearance of diurnal cycle.

\acknowledgments
Selected model output and code for the model setup and plots are available at \url{https://github.com/wanyingkang/shortwave-absorptive-atmosphere.git}. The Isca model source code can be downloaded from \url{https://github.com/ExeClim/Isca}. R.W. acknowledges support from NASA Habitable Worlds grant NNX16AR86G. W.K. was supported by the NSF Climate and Large-Scale Dynamics program, grant AGS-1826635, and by the NASA Habitable Worlds program grant FP062796-A/NNX16AR85G.
We would also like to acknowledge high-performance computing support from Cheyenne provided by NCAR's Computational and Information Systems Laboratory, sponsored by the National Science Foundation.

\end{document}